\documentclass[a4paper,fleqn]{cas-sc}

\usepackage[authoryear,longnamesfirst]{natbib}
\usepackage{subcaption}
\usepackage[ascii]{inputenc}
\usepackage{graphicx}
\usepackage{subcaption}
 \usepackage{setspace}
\usepackage{textcomp}     
\usepackage{url}
\usepackage[]{amsmath}
\usepackage{amsmath}
\usepackage{nccmath}
\usepackage{lineno}

\newcommand{\an}{{\textquotesingle Ayl\'{o}\textquotesingle chaxnim }}
\newcommand{\anns}{{\textquotesingle Ayl\'{o}\textquotesingle chaxnim}}

\newcommand{\ans}{{\textquotesingle Ayl\'{o}\textquotesingle chaxnims }}
\newcommand{\ansns}{{\textquotesingle Ayl\'{o}\textquotesingle chaxnims}}
\newcommand{\nan}{{(594913) \textquotesingle Ayl\'{o}\textquotesingle chaxnim }}
\newcommand{\nanns}{{(594913) \textquotesingle Ayl\'{o}\textquotesingle chaxnim}}

\begin{document}
\let\WriteBookmarks\relax
\def\floatpagepagefraction{1}
\def\textpagefraction{.001}

\shorttitle{Preliminary estimates of the ZTF \an population completeness}

\shortauthors{Bolin et al.}  

  \title[mode = title]{Preliminary estimates of the Zwicky Transient Facility \an asteroid population completeness}
\author[]{B. T. Bolin$^{a,b,c,*,**}$}[orcid=0000-0002-4950-6323]
\author[4]{T. Ahumada}[]
\author[5]{P. van Dokkum}[]
\author[6]{C. Fremling}[]
\author[7]{K. K. Hardegree-Ullman}[]
\cormark[3]
\author[6]{J. N. Purdum}[]
\author[8]{E. Serabyn}[]
\author[9]{J. Southworth}[]

\address[1]{Goddard Space Flight Center, 8800 Greenbelt Road, Greenbelt, MD 20771, USA}
\address[2]{Division of Physics, Mathematics and Astronomy, California Institute of Technology, Pasadena, CA 91125, USA}
\address[3]{Infrared Processing and Analysis Center, California Institute of Technology, Pasadena, CA 91125, USA}
\address[4]{Department of Astronomy, University of Maryland, College Park, MD 20740, USA}
\address[5]{Department of Astronomy, Yale University, New Haven, CT 06511, USA}
\address[6]{Caltech Optical Observatory, California Institute of Technology, Pasadena, CA 91125, USA}
\address[7]{Steward Observatory, University of Arizona, Tucson, AZ 85721, USA}
\address[8]{Jet Propulsion Laboratory, California Institute of Technology, Pasadena, CA 91109, USA}
\address[9]{Astrophysical Research Institute, Liverpool John Moores University, Liverpool, L2 2QP, UK}
\cortext[cor1]{Corresponding author: bolin.astro@gmail.com}
\cortext[cor2]{NASA Postdoctoral Program Fellow.} 
\cortext[cor3]{Visiting astronomer, Cerro Tololo Inter-American Observatory at NSF's NOIRLab, which is managed by the Association of Universities for Research in Astronomy (AURA) under a cooperative agreement with the National Science Foundation.} 

\begin{abstract}[S U M M A R Y]
Near-Earth asteroids (NEAs) are organized into five main classes: Amor, Apollo, Aten, Atira and \anns. Asteroids belonging to the \an class are located entirely within the orbit of Venus making them difficult to detect by ground-based observatories. The first-known asteroid of this class, \nanns, was discovered by the Zwicky Transient Facility (ZTF) in 2020 January during a twilight search for asteroids at small solar elongations that ran between September 2019 and January 2020. Due to its large diameter of $\sim$2 km, the discovery of \nan is surprising because contemporary NEA population models predict a scarcity of asteroids of this size located inside the orbit of Venus. To compare the discovery of \nan by ZTF with the predictions of NEA population models, we estimated the ZTF survey completeness at detecting \an asteroids and the number of \an asteroids expected to have been discovered by simulating observations of synthetic \an asteroids. We find that the \an population completeness of the survey is $\sim$18$\%$ and there is only a 5$\%$ probability that a single \an asteroid would have been discovered. Given the small chance for \nan to have been discovered, its presence is either a statistical fluke or it implies that asteroid population models may need to be revised.
\end{abstract}

\maketitle
\section{Introduction}
There are presently over 30,000 known near-Earth asteroids\footnote{\url{https://cneos.jpl.nasa.gov/stats/totals.html}}, defined as asteroids that have a perihelion distance, $q$<1.3 au \cite{Binzel2015}. The observed population of near-Earth asteroids (NEAs) is a function of both the underlying population and discovery biases \cite{Jedicke2015}. Contemporary models which describe the underlying asteroid population predict that there should be $\sim$800,000 NEAs with absolute magnitude, $H$, brighter than 25 \cite{Granvik2018,Morbidelli2020}. NEAs are divided into five distinct classes: Amors, Apollos, Atens, Atiras and \ansns. Sub-clasess exist within these main NEA classes, e.g., \cite{Granvik2013ab,Jedicke2018, Bolin2020CD3}, however, we will describe only NEAs from these main classes. Amor asteroids have semi-major axes, $a$, greater than 1 au and 1.017 au $<$$q$$<$ 1.3 au, Apollo asteroids also have $a$$>$ 1 au but have $q$$<$1.017 au, Aten asteroids have $a$$<$1 au and aphelia, $Q$, greater than 0.983 au, Atira asteroids have $Q$$<$0.983 au, located entirely within the orbit of the Earth, and \an asteroids have $Q$$<$0.718 au located entirely with the orbit of Venus \cite{Binzel2015, Bolin2022}. We note that the class of inner-Venus asteroids was provisionally called ``Vatira'', however, following a long-established tradition, this class is now named after the first-discovered object, \nanns. The majority of the NEA population with $H$$<$25 consists of Apollo asteroids comprising $\sim$55$\%$ of the total NEA population with the next largest group being the Amors comprising of $\sim$40$\%$ of the NEAs \cite{Granvik2018}. The remaining 5$\%$ of asteroids described by the NEA population model consists of the three remaining Aten, Atira and \an groups with \ans being the most scarce consisting of only $\sim$0.3$\%$ of the total model NEA population. 

There are presently $\sim$30 known Atira asteroids\footnote{\url{https://minorplanetcenter.net/iau/lists/MPLists.html}}, most of which have been discovered by survey telescopes designed to detect NEAs such as the Catalina Sky Survey \cite{Zavodny2008,Granvik2016} and the Panoramic Survey Telescope and Rapid Response System (Pan-STARRS) survey \cite{Denneau2013,Chambers2016}. A number of Atira asteroids have been discovered with bespoke surveys designed specifically to detect asteroids on orbits interior to the orbit of the Earth using the University of Hawai`i 2.2-meter telescope \cite{Whitely1998,Tholen1998} and the V\'{i}ctor M. Blanco 4-meter telescope \cite{Pokorny2020, Sheppard2022}. Although these surveys have been adept at finding Atira asteroids, before 2020, an \an asteroid had not yet been discovered.

The Zwicky Transient Facility (ZTF) is a northern hemisphere transient survey using the Palomar 48-inch telescope \cite{Bellm2019, Graham2019}. ZTF covered portions near the Sun during evening and morning twilight enabling it to detect asteroids at small solar elongation angles \cite{Bolin2022}. ZTF covered $\sim$40,000 sq. degrees of sky within 35-55 degrees of the Sun during evening and morning Twilight between 2019 September and 2020 January. The first known asteroid located entirely within the orbit of Venus, \nanns, was discovered by ZTF in the evening twilight sky at $V$$\sim$18.2 while only $\sim$40 degrees from the Sun on 2020 January 4 \cite{Bolin2020MPEC, Bolin2022}. The maximum solar elongation distance of \an asteroids is 47 degrees thereby making the 35-55 degrees solar elongation coverage by ZTF ideal for their detection.

Current estimates of the number of \an asteroids in NEA population are based on asteroid survey data that predate the discovery of \nan \cite{Granvik2018}. In addition, the NEA model assumes that NEAs originate from the Main Asteroid Belt \cite{Granvik2017}. Limits on the \an population have been based on the non-detection of \an asteroids in other recent near-Sun surveys \cite{Sheppard2022}. Thus the discovery of \nan by ZTF provides an opportunity to compare the NEA model's prediction of the population of \an asteroids with asteroid survey data that includes the first known example of this class of asteroids. A preliminary estimate of the comparison between the NEA model and the discovery of \an was provided in \cite{Bolin2022} but did not take into account the coverage of the ZTF survey. A description of the analysis of the completeness of the ZTF survey for the detection of \an asteroids and a comparison with the NEA model is described in the following text. 

\section{Methods}
We estimated the number of \ans expected to have been discovered by the Twilight Survey by simulating ZTF observations of synthetic \ansns. We generated the synthetic \an population from the NEA model \citep[][]{Granvik2018} using their medium resolution simulations with a semi-major axis, $\Delta a$ resolution element of 0.05~au, an eccentricity resolution element, $\Delta e$ of 0.02, and inclination resolution element, $\Delta i$ of 2.0$^\circ$ and an absolute magnitude resolution element, $\Delta H$, of 0.25~mag oversampling a single version of the NEA model by a factor of 1,000, producing $\sim$3.0$\times10^3$ inner-Venus objects with 15$<$$H$$<$18. We note that generating a model oversampled by a factor of 1,000 may result in under-estimated uncertainties compared to generating 1,000 separate models and combining them. We used synthetic \ans with 15 $<$ $H$ $<$ 18 because this brackets the absolute magnitude value of \nan of $H$ = 16.2$\pm$0.8~mag taken from from the JPL Small Body Database\footnote{\url{https://ssd.jpl.nasa.gov/tools/sbdb\_lookup.html\#/?sstr=594913}}. The $a$, $e$, $i$, $H$ and $Q$ distributions of our synthetically generated \ans are plotted in Fig.~1 (A-E). To simulate the observations of our synthetic inner-Venus asteroids by ZTF, we used the complete list of the Twilight Survey telescope pointings between 2019 September 20 and 2020 January 30 \cite{Bolin2022} using the apparent brightness of the synthetic \ans in a survey simulation as seen in Fig.~S1 (A).

Synthetic observations of \ans were generated using \textsc{Near-Earth Object Survey Simulation} \cite{Naidu2017} and the \an population described by the NEA model as input test particles into the simulation \cite{Granvik2018}. We refined the synthetic \an observations by calculating the efficiency for each synthetic \an detection in each field. The per object per field efficiency is estimated by comparing actual detections of known moving objects serendipitously observed in the Twilight Survey fields with the predicted number of known objects detected in the fields. The average per field detection efficiency as a function of visible $V$ magnitude is presented in Fig.~1 (F). The average per field detection efficiency calculation was based on the detection of known objects in Twilight Survey fields between September 2019 and January 2020. The function describing the efficiency as a function of $V$ magnitude is $\epsilon (V)$ and defined as
\begin{align}
&&\epsilon (V) = \epsilon_0\left[1 + \mathrm{exp}\left(\frac{V-V_{\mathrm{lim}}}{V_{\mathrm{width}}}\right)\right]^{-1}
\end{align}
where $\epsilon_0$ represents the maximum possible efficiency for detecting moving objects, $V_{\mathrm{lim}}$ is the effective limiting magnitude of the survey where the efficiency drops to half for detecting moving objects, and $V_{\mathrm{width}}$ is the width of the drop in the detection efficiency of faint moving objects \cite{Jedicke2016}. The average per-field efficiency is interpolated using Eq.~(1) and $\epsilon_0$ = 0.87, $V_{\mathrm{lim}}$ = 20.60~mag and $V_{\mathrm{width}}$ = 0.74, plotted in Fig.~1 (F). The value of $\epsilon_0$ = 0.87 is close to the fraction of area covered by camera detectors in the ZTF image plane \cite{Bellm2019} suggesting that the limiting factor in detecting bright moving objects by ZTF is the detector layout rather than the ZTF processing pipeline \cite{Masci2019}. An $\epsilon (V)$ = 0.87 is assumed for bright object detections with V$<$17.5. A histogram of the exposure-based limiting magnitudes is presented in Fig.~S1 (C); these are the 5-$\sigma$ limiting values in the Twilight survey pointing with no assumption or correction regarding the efficiency of asteroid detections. Therefore, the raw 5-$\sigma$ limiting exposure-based survey magnitudes may actually correspond to higher efficiency levels on average.

In addition to the limiting magnitude, trailing losses due to the sky plane motion of the \ans could further decrease the efficiency calculated with Eq.~(1) \cite{Jedicke2016}. However, as seen in Fig.~S1 (B), the vast majority of synthetic \an detections have a sky plane rate of motion of 240 arcseconds per hour or slower which does not result in substantial trailing of the detections in a single 30 s Twilight Survey exposure given the typical 2 arcseconds seeing at the Palomar observing site. Therefore, we do not expect the detection of \an asteroids to be affected by trailing losses.

The efficiency per \an asteroid, $j$, per observing session, $n$, $\epsilon_{j,n} (V_{j,n})$ is given by its per session field visible magnitude, $V_{j,n}$ using Eq.~(1). If a synthetic \an is not seen $>$3 times, $\epsilon_{j,n} (V_{j,n}) = 0$. The vast majority of synthetic \ans generated with 15 $<$$H$$<$18 have $V$ magnitude $<$20 as seen in Fig.~S1 (A) resulting in the majority of \an detections located in Twilight Survey fields having $\epsilon_{j,n} (V_{j,n})$$>$70$\%$.

The probability of detecting a single Synthetic \anns, $j$, $p_j$,over the total $n_{\mathrm{max}}$ Twilight Survey sessions is given by 
\begin{align}
&&p_j = 1 - \prod^{n_{\mathrm{max}}}_{n=0} \left [1 - \epsilon_{j,n} (V_{j,n})\right ]
\end{align}
where $n_{max}$ = 89 corresponding to 90 individual Twilight Survey sessions. The number of detected synthetic \ans is weighted per synthetic \an $a$, $e$, $i$ and $H$ bin by each detected object's $p_j$. The completeness per synthetic \an $a$, $e$, $i$ and $H$ bin is calculated by dividing the weighted number of detected synthetic \ans by the total number of synthetic \ans generated from the NEA model per $a$, $e$, $i$ and $H$ bin. 

\section{Results}
The survey simulator output shows synthetic inner-Venus asteroids detections that we use to estimate the completeness of Twilight Survey at detecting inner-Venus asteroids. The completeness of \ans detected per $a$, $e$, $i$ and $H$ bin is plotted in Fig.~1 (A-E). The peak at $\sim$0.5 au in the $a$ distribution of \an asteroids may be due to the fact that asteroids with this semi-major axis are more distant from the orbits of Venus and Mercury and thus will have fewer encounters with these planets affecting their orbits. Averaging over the complete $a$, $e$, $i$ and $H$ distribution of the synthetic \an population results in an completeness of $\sim$0.15 and a completeness of $\sim$0.18 when considering only $H$ bins covering \anns's $H$ magnitude of $H$ = 16.2$\pm$0.8. The completeness as a function of $H$ is $\sim$0.18 for most bins of $H$ in Fig.~1 (D) due to the fact that the majority of synthetic asteroids have V$<$19.5 as seen in Fig.~S1(A) which all have a detection efficiency of $\sim$80$\%$ as seen in Fig.~1 (F). We repeated the completeness calculation use a variety of slope parameters representative of the different Q, S, C, D and X taxonomic types \cite{Pravec2012,Veres2015} and found that the different slope parameter values did not substantially alter the completeness result. As discussed above, assuming a slightly larger phase parameter for \an to estimate its absolute magnitude only slightly increases the number of \ans predicted by the NEA model and thus only increases the expected number of inner-Venus objects detected in the twilight survey by $\sim$1$\%$.

The number of \ans in the NEA model brighter than \nanns's nominal value of $H$ = 16.2 is 0.25 \cite{Granvik2018}. As described in \cite{Bolin2022}, because of the 1-$\sigma$ uncertainty of 0.8 on the $H$ of \nanns, the number of asteroids brighter than the 1-$\sigma$ $H$ lower bound of 15.4 and the 1-$\sigma$ $H$ upper bound of 17.0 is uneven, with the lower bound corresponding to 0.05 objects, and the upper bound corresponding to 0.7 objects. The number of objects brighter than the 1-$\sigma$ upper and 1-$\sigma$ lower bound of $H$ $<$ 16.2$\pm$0.8 is thus 0.25$\pm$$^{0.45}_{0.20}$. The NEA model predicts that there should be $\sim$1 \an asteroid with $H$ = 17.5. Thus \nan is $\sim$1.5 magnitudes brighter or $\sim$2 times larger in size than the largest \an asteroid that the NEA model predicts should exist. There is inherent uncertainty in the number of asteroids predicted by the NEA model as a function of $H$ \cite{Granvik2018}, however, we assume that the main contributor to the uncertainty in the number of asteroids is the 0.8 magnitude uncertainty on the value of $H$ for \nanns. 

Combining the number of \an asteroids with $H$ $<$ 16.2$\pm$0.8 of 0.25$\pm$$^{0.45}_{0.20}$ with the $0.18\pm0.02$ completeness estimate from above, we expect 0.05$\pm$$^{0.09}_{0.04}$ \ans to have been discovered during the survey. The completeness drops severely with smaller values of $Q$ than the $Q$$\sim$0.65~au of \an as seen in (Fig.~1E) suggesting that its detection was barely within the capabilities of the Twilight Survey.

\section{Discussion and conclusion}
Given the $\sim$0.18 completeness for detecting \an asteroids and the few number of km-scale \an asteroids predicted by the NEA model, the detection of a one km-scale \an by the Twilight Survey is suprising. The NEA model predicts there should be $\sim$1 \an object with H$<$18.5. Assuming a survey completeness of $\sim$0.18 for 5 H<18.5 \an objects, ZTF should have found $\sim$1 \an object with H<18.5. (594913) \an has H = 16.2 +/-0.8, therefore, its discovery  is inconsistent with the Granvik NEO model at the 3-sigma level. However, this only includes the uncertainty on H and there may be other uncertainties that are not being taken into account.

If not a fluke, a possible explanation for the discovery of \nan despite its low chance of discovery is that it could have originated from a larger population of \ans than predicted by contemporary NEA models. A potential explanation for a larger population of \ans could be the location of asteroids in stability regions closer to the Sun, such as those located inside the orbit of Mercury at $\sim$0.1-0.2~au \cite{Evans1999}. Kilometer-scale asteroids located here could have previously gone undetected \cite{Steffl2013}, and would remain stable and intact over the age of the Solar System \cite{Vokrouhlick2000Vul, Bottke2005b, Shannon2015}. Comparison of a possible source of asteroids close to the Sun can be made with the apparent detection of exozodiacal dust within 0.1 au as seen in near-infrared interferometric observations of other stars \cite{Absil2006,Ertel2014}, therefore, the regions of stability inside Mercury's orbit bear further dynamical and observational investigation.

\bibliographystyle{Science}
\bibliography{scibib}

\begin{thebibliography}{10}

\bibitem{Binzel2015}
R.~P. {Binzel}, V.~{Reddy}, T.~L. {Dunn}, {\it {The Near-Earth Object
  Population: Connections to Comets, Main-Belt Asteroids, and Meteorites}\/}
  (2015), pp. 243--256.

\bibitem{Jedicke2015}
R.~{Jedicke}, {\it et~al.\/}, {\it Asteroids IV\/} pp. 795--813 (2015).

\bibitem{Granvik2018}
M.~{Granvik}, {\it et~al.\/}, {\it \icarus\/} {\bf 312}, 181 (2018).

\bibitem{Morbidelli2020}
A.~{Morbidelli}, {\it et~al.\/}, {\it \icarus\/} {\bf 340}, 113631 (2020).

\bibitem{Granvik2013ab}
M.~{Granvik}, R.~{Jedicke}, B.~{Bolin}, M.~{Chyba}, G.~{Patterson}, {\it
  {Earth's Temporarily-Captured Natural Satellites - The First Step towards
  Utilization of Asteroid Resources}\/} (2013), pp. 151--167.

\bibitem{Jedicke2018}
R.~{Jedicke}, {\it et~al.\/}, {\it Frontiers in Astronomy and Space Sciences\/}
  {\bf 5}, 13 (2018).

\bibitem{Bolin2020CD3}
B.~T. {Bolin}, {\it et~al.\/}, {\it \apjl\/} {\bf 900}, L45 (2020).

\bibitem{Bolin2022}
B.~T. {Bolin}, {\it et~al.\/}, {\it \mnras\/} {\bf 517}, L49 (2022).

\bibitem{Zavodny2008}
M.~{Zavodny}, R.~{Jedicke}, E.~C. {Beshore}, F.~{Bernardi}, S.~{Larson}, {\it
  Icarus\/} {\bf 198}, 284 (2008).

\bibitem{Granvik2016}
M.~{Granvik}, {\it et~al.\/}, {\it \nat\/} {\bf 530}, 303 (2016).

\bibitem{Denneau2013}
L.~{Denneau}, {\it et~al.\/}, {\it \pasp\/} {\bf 125}, 357 (2013).

\bibitem{Chambers2016}
K.~C. {Chambers}, {\it et~al.\/}, {\it ArXiv e-prints\/}  (2016).

\bibitem{Whitely1998}
R.~J. {Whiteley}, D.~J. {Tholen}, {\it \icarus\/} {\bf 136}, 154 (1998).

\bibitem{Tholen1998}
D.~J. {Tholen}, R.~J. {Whiteley}, {\it AAS/Division for Planetary Sciences
  Meeting Abstracts \#30\/} (1998), vol.~30 of {\it AAS/Division for Planetary
  Sciences Meeting Abstracts\/}, p. 16.04.

\bibitem{Pokorny2020}
P.~{Pokorn{\'y}}, M.~J. {Kuchner}, S.~S. {Sheppard}, {\it \psj\/} {\bf 1}, 47
  (2020).

\bibitem{Sheppard2022}
S.~S. {Sheppard}, {\it et~al.\/}, {\it \aj\/} {\bf 164}, 168 (2022).

\bibitem{Bellm2019}
E.~C. {Bellm}, {\it et~al.\/}, {\it \pasp\/} {\bf 131}, 018002 (2019).

\bibitem{Graham2019}
M.~J. {Graham}, {\it et~al.\/}, {\it \pasp\/} {\bf 131}, 078001 (2019).

\bibitem{Bolin2020MPEC}
B.~T. {Bolin}, {\it et~al.\/}, {\it Minor Planet Electronic Circulars\/} {\bf
  2020-A99} (2020).

\bibitem{Granvik2017}
M.~{Granvik}, {\it et~al.\/}, {\it \aap\/} {\bf 598}, A52 (2017).

\bibitem{Naidu2017}
S.~P. {Naidu}, S.~R. {Chesley}, D.~{Farnocchia}, {Near-Earth Object Survey
  Simulation Software},
  \url{https://github.com/AsteroidSurveySimulator/objectsInField}. Accessed:
  2021 August 26.

\bibitem{Jedicke2016}
R.~{Jedicke}, B.~{Bolin}, M.~{Granvik}, E.~{Beshore}, {\it \icarus\/} {\bf
  266}, 173 (2016).

\bibitem{Masci2019}
F.~J. {Masci}, {\it et~al.\/}, {\it \pasp\/} {\bf 131}, 018003 (2019).

\bibitem{Pravec2012}
P.~{Pravec}, A.~W. {Harris}, P.~{Ku{\v s}nir{\'a}k}, A.~{Gal{\'a}d},
  K.~{Hornoch}, {\it \icarus\/} {\bf 221}, 365 (2012).

\bibitem{Veres2015}
P.~{Vere{\v s}}, {\it et~al.\/}, {\it \icarus\/} {\bf 261}, 34 (2015).

\bibitem{Evans1999}
N.~W. {Evans}, S.~{Tabachnik}, {\it \nat\/} {\bf 399}, 41 (1999).

\bibitem{Steffl2013}
A.~J. {Steffl}, N.~J. {Cunningham}, A.~B. {Shinn}, D.~D. {Durda}, S.~A.
  {Stern}, {\it \icarus\/} {\bf 223}, 48 (2013).

\bibitem{Vokrouhlick2000Vul}
D.~{Vokrouhlick{\'y}}, P.~{Farinella}, W.~F. {Bottke}, {\it \icarus\/} {\bf
  148}, 147 (2000).

\bibitem{Bottke2005b}
W.~F. {Bottke}, {\it et~al.\/}, {\it \icarus\/} {\bf 179}, 63 (2005).

\bibitem{Shannon2015}
A.~{Shannon}, A.~P. {Jackson}, D.~{Veras}, M.~{Wyatt}, {\it \mnras\/} {\bf
  446}, 2059 (2015).

\bibitem{Absil2006}
O.~{Absil}, {\it et~al.\/}, {\it \aap\/} {\bf 452}, 237 (2006).

\bibitem{Ertel2014}
S.~{Ertel}, {\it et~al.\/}, {\it \aap\/} {\bf 570}, A128 (2014).

\end{thebibliography}

\section*{Acknowledgements}

\noindent The authors would like to acknowledge Frank Masci for help with operating the software for identifying asteroids in ZTF data, helping with the completeness estimate, and providing helpful comments on the manuscript. The authors would also like to thankAlessandro Morbidelli for the useful discussion and for providing the synthetic asteroid population. C.F.~acknowledges support from the Heising-Simons Foundation (grant $\#$2018-0907). Part of this work was carried out at the Jet Propulsion Laboratory, California Institute of Technology, under contract with NASA 80NM0018D0004. Based on observations obtained with the Samuel Oschin 48-inch Telescope at the Palomar Observatory as part of the Zwicky Transient Facility project. ZTF is supported by the National Science Foundation under Grant No. AST-1440341 and a collaboration including Caltech, IPAC, the Weizmann Institute for Science, the Oskar Klein Center at Stockholm University, the University of Maryland, the University of Washington, Deutsches Elektronen-Synchrotron and Humboldt University, Los Alamos National Laboratories, the TANGO Consortium of Taiwan, the University of Wisconsin at Milwaukee, and Lawrence Berkeley National Laboratories. Operations are conducted by COO, IPAC, and UW.



\subsection*{Software}

\noindent The \textsc{Near-Earth Object Survey Simulation} software \cite{Naidu2017} are publicly available.

\newpage
\clearpage
\begin{figure}[ht]
\begin{subfigure}[b]{0.45\linewidth}
\centering
\hspace{-1.0cm}
\includegraphics[width=1.0\linewidth]{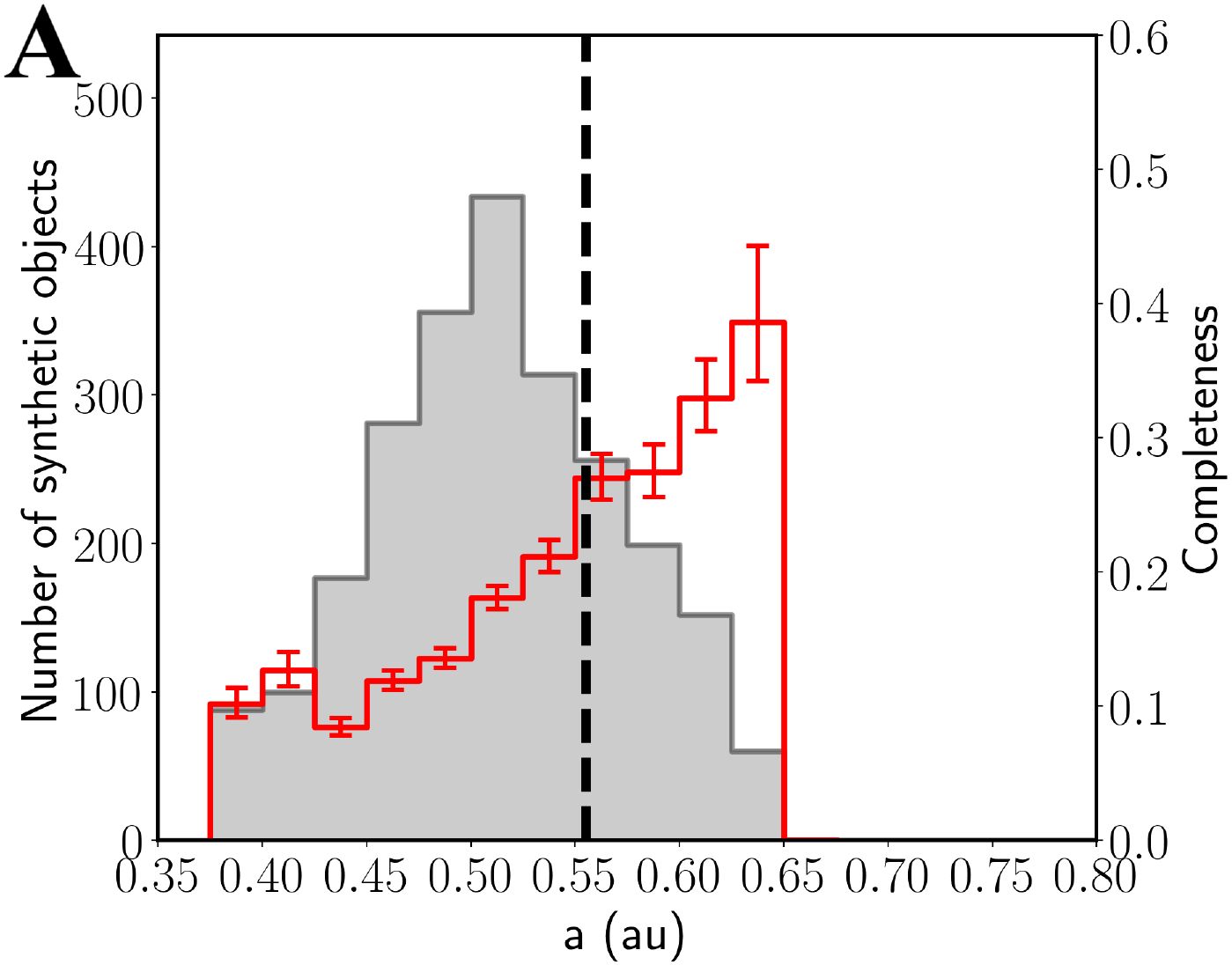}
\label{fig7:a}
\vspace{4ex}
\end{subfigure}
\begin{subfigure}[b]{0.45\linewidth}
\centering
\hspace{-1.0cm}
\includegraphics[width=1.0\linewidth]{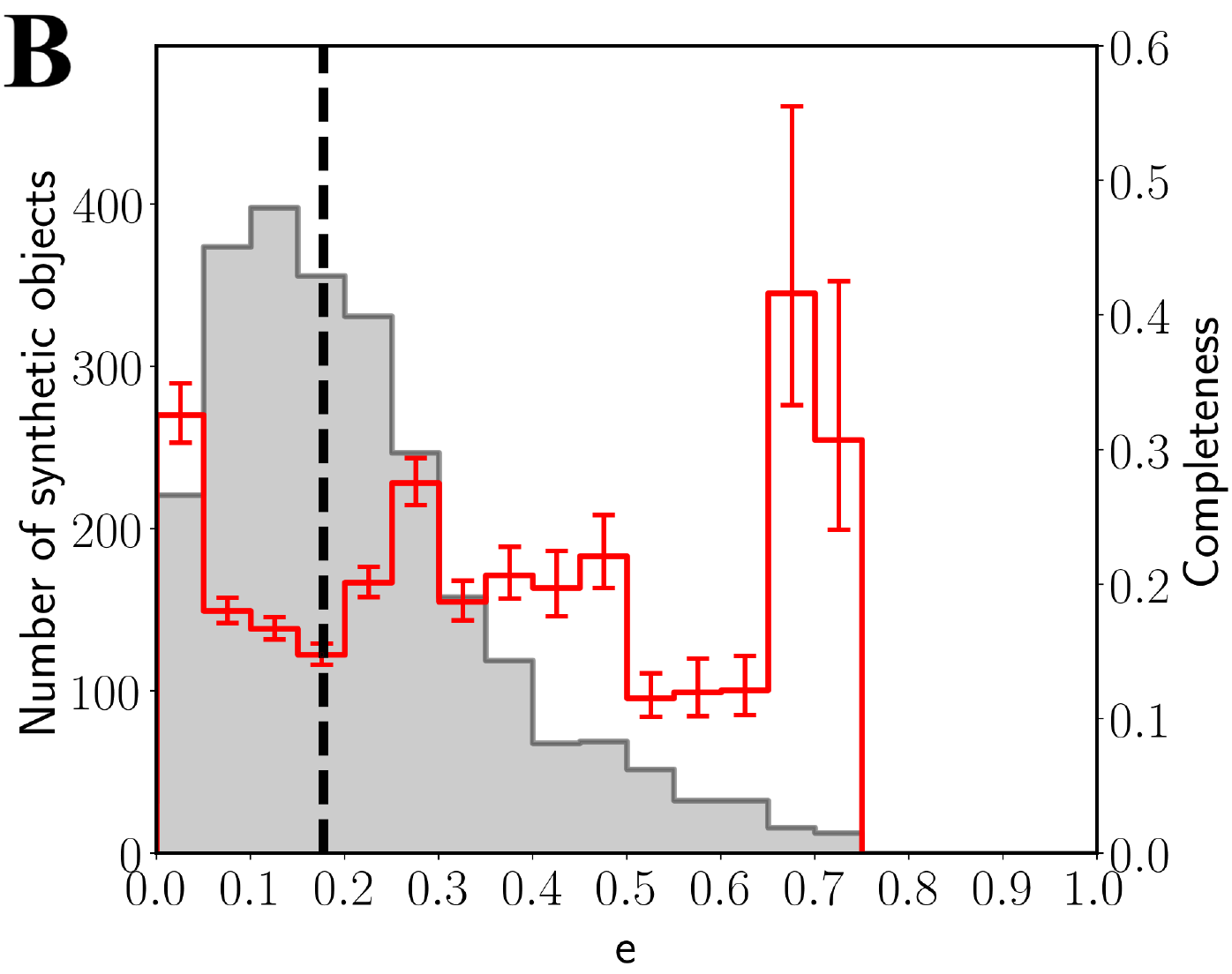}
\label{fig7:b}
\vspace{4ex}
\end{subfigure}
\begin{subfigure}[b]{0.45\linewidth}
\centering
\hspace{-1.0cm}
\includegraphics[width=1.0\linewidth]{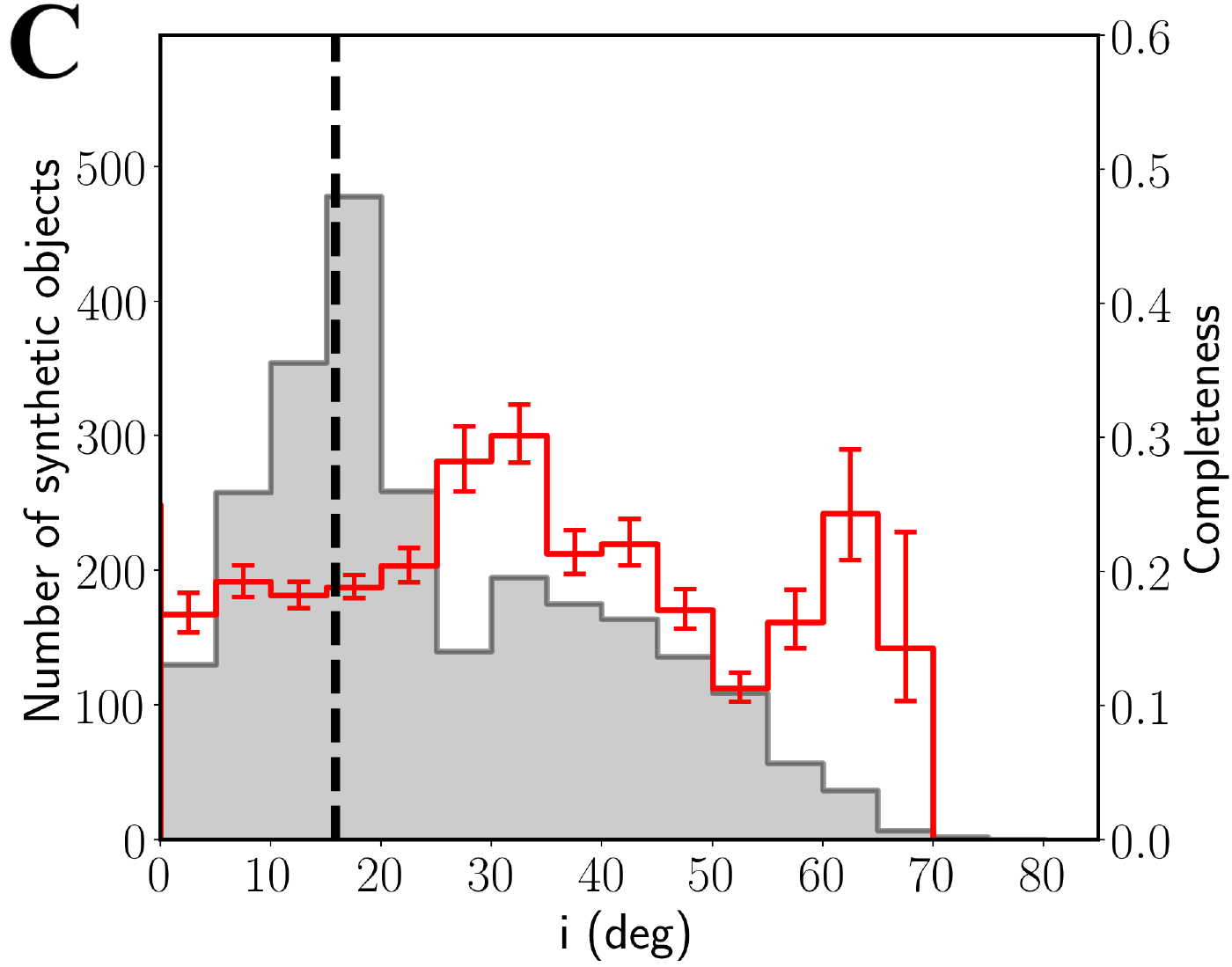}
\label{fisg7:c}
\end{subfigure}
\begin{subfigure}[b]{0.45\linewidth}
\centering
\hspace{-1.0cm}
\includegraphics[width=1.0\linewidth]{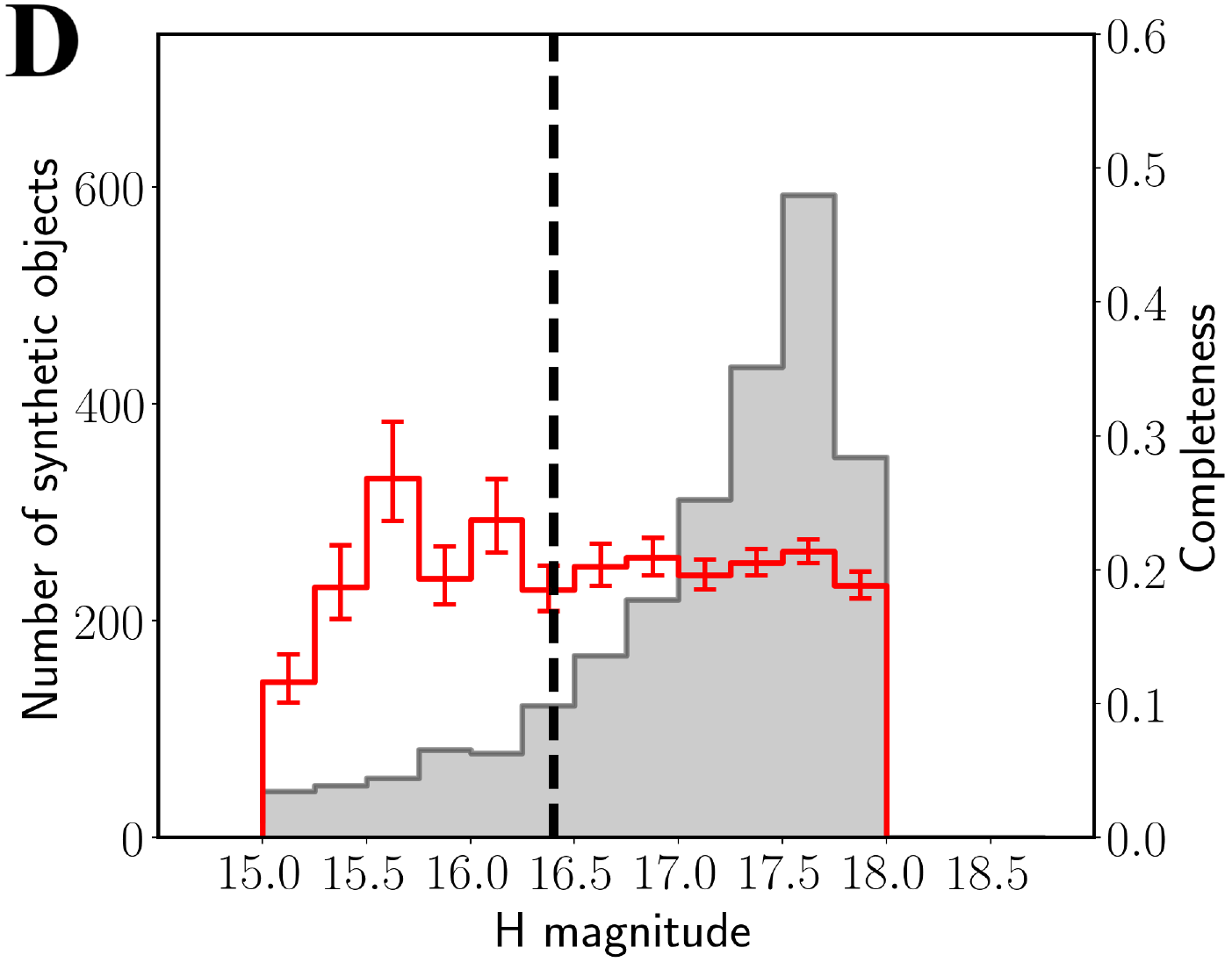}
\label{fig7:d}
\end{subfigure}
\begin{subfigure}[b]{0.45\linewidth}
\centering
\hspace{-3.5cm}
\includegraphics[width=1.0\linewidth]{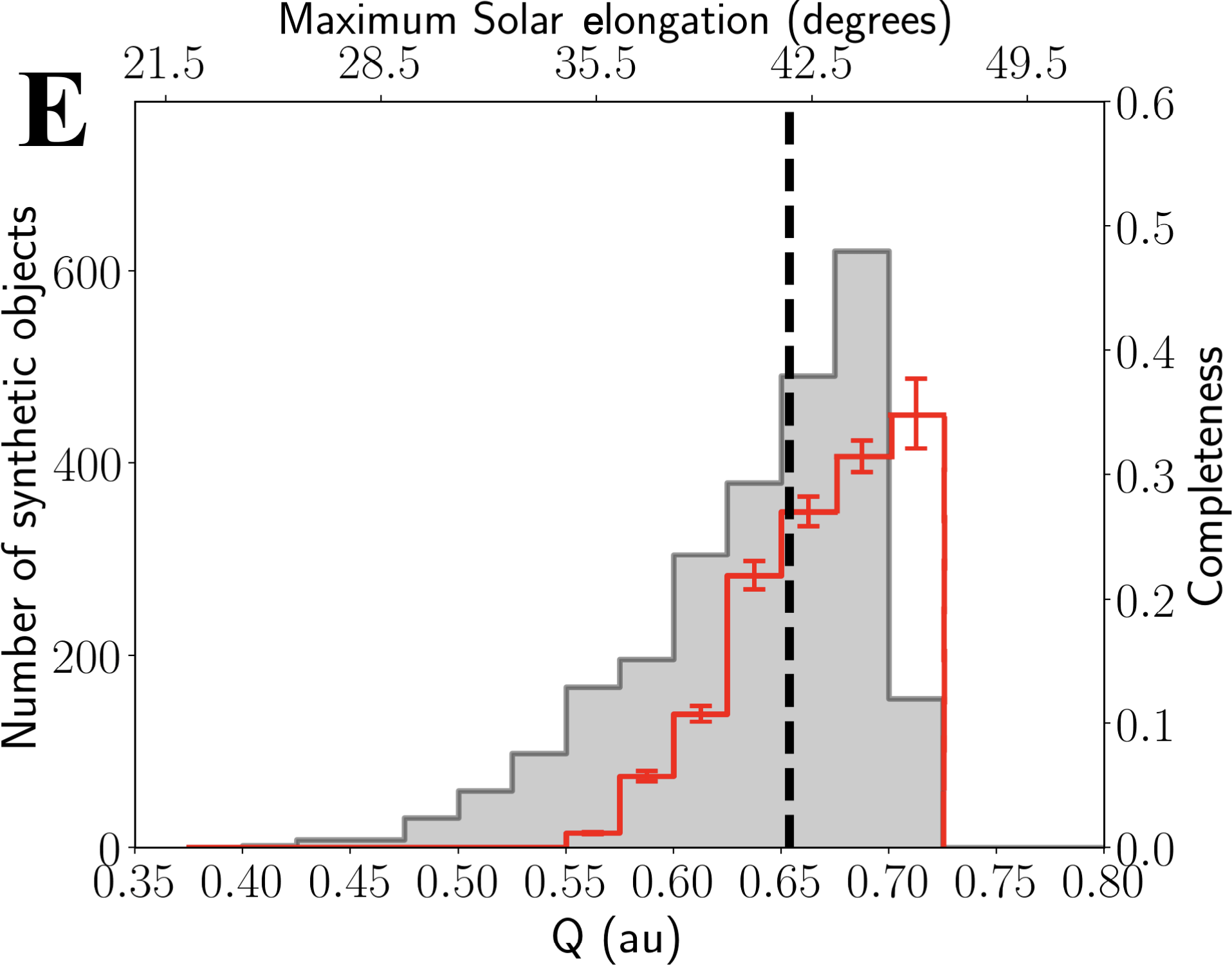}
\label{fig7:e}
\end{subfigure}
\begin{subfigure}[b]{.38\linewidth}
\centering
\hspace{-2.75cm}
\includegraphics[width=1.0\linewidth]{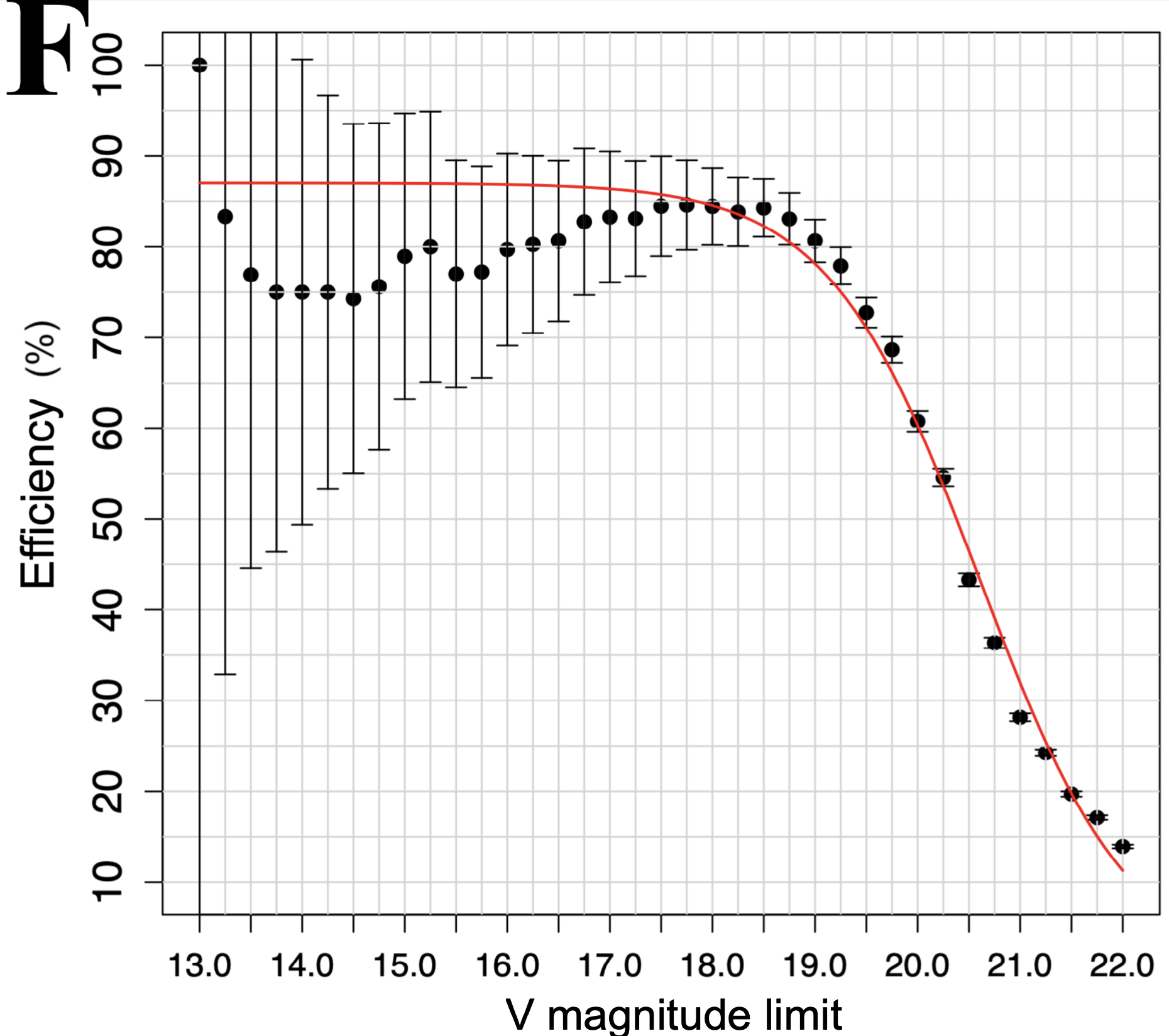}
\label{fig7:f}
\end{subfigure}
\caption{Orbital distribution and completeness of synthetic inner-Venus asteroids. Comparison of the (A) semi-major axis, $a$, (B) eccentricity, $e$, (C) inclination $i$,  (D) absolute magnitude $H$ and (E) Aphelion, $Q$, distributions of the synthetic inner-Venus asteroids generated from the NEA model \cite{Granvik2018} (grey histograms) with the completeness calculated in the survey simulation (red histogram). The lower level of the last bin in panel (D) is due to histogram bin aliasing. The 1-$\sigma$ error bars on the completeness are determined assuming Poisson statistics. The vertical dashed black line indicates the value of each orbital element for \nanns. The absolute magnitude range of 15 to 18 corresponds to asteroids in the size range of $\sim$1-3 km assuming a 20$\%$ albedo. The synthetic objects have been generated  from a single version of the NEA model and oversampled by a factor of 1,000. (F) The average per field detection efficiency of asteroids in the Twilight survey as a function of $V$ magnitude. The detection efficiency calculation was based on the detection of known objects in Twilight Survey fields. The 1-$\sigma$ error bars are determined assuming Poisson statistics though Poisson statistics may be less applicable for brighter asteroids. Eq.~(1) is plotted as a red line with $\epsilon_0$ = 0.87, $V_{\mathrm{lim}}$ = 20.60, $V_{\mathrm{width}}$ = 0.74.}
\end{figure}

\clearpage
\newpage
\renewcommand{\thefigure}{S\arabic{figure}}
\setcounter{figure}{0}
\renewcommand{\thetable}{S\arabic{table}}s
\renewcommand{\theequation}{S\arabic{equation}}
\renewcommand{\thesection}{S\arabic{section}}
\setcounter{section}{0}

\newpage
\section*{Supplementary Material}
\newpage
\begin{figure}
\centering 
\includegraphics[width=0.465\linewidth]{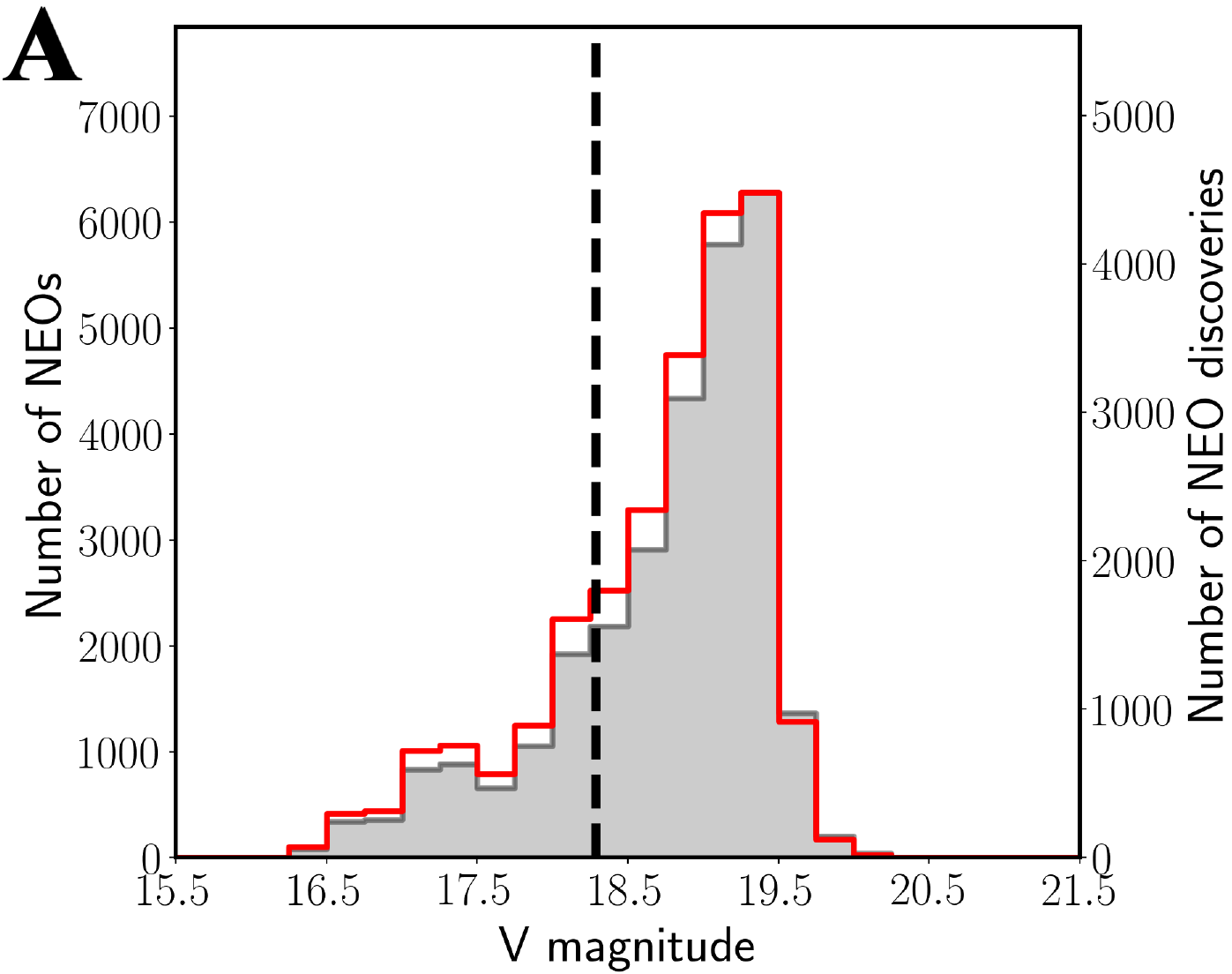}
\includegraphics[width=0.465\linewidth]{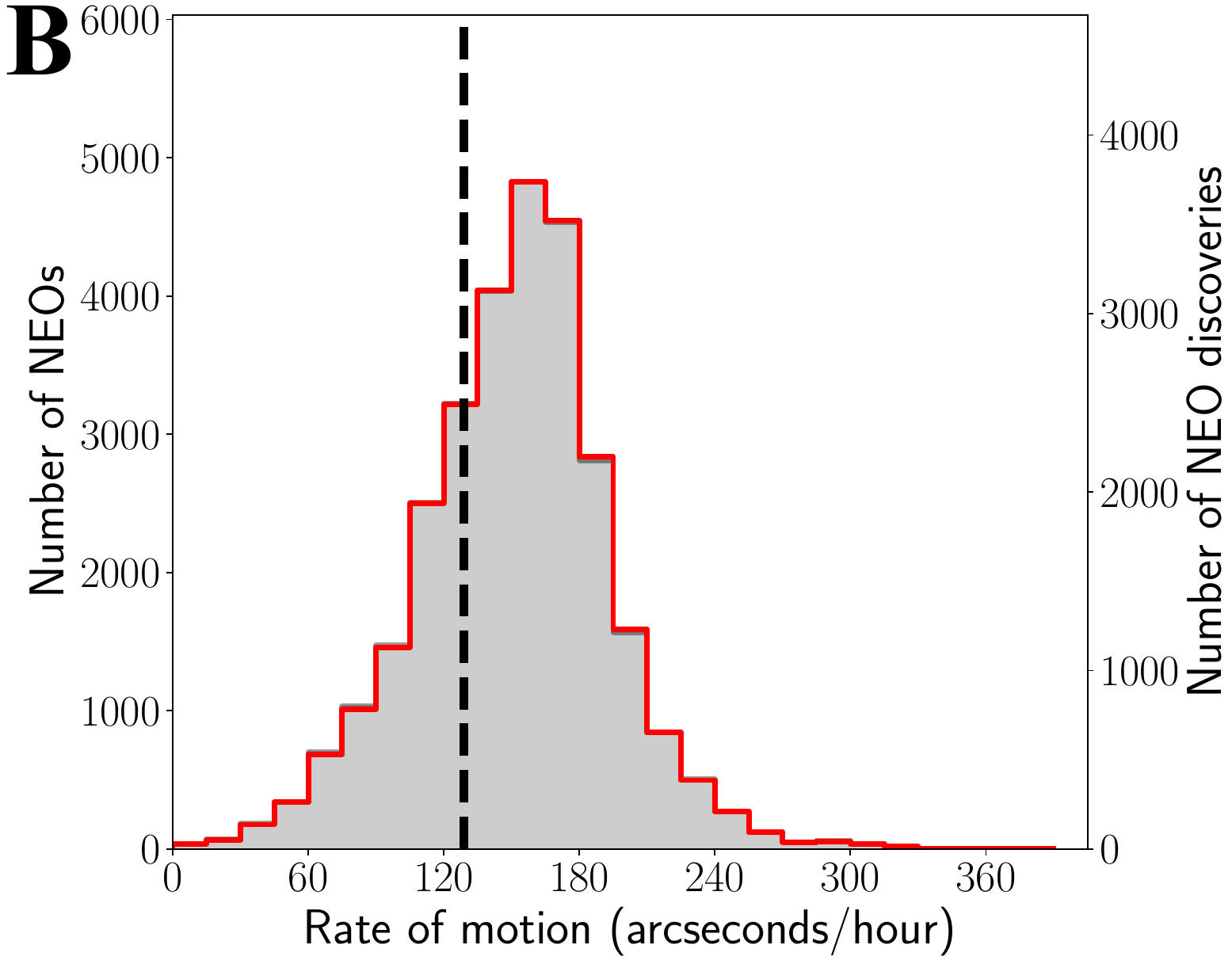}
\includegraphics[width=0.465\linewidth]{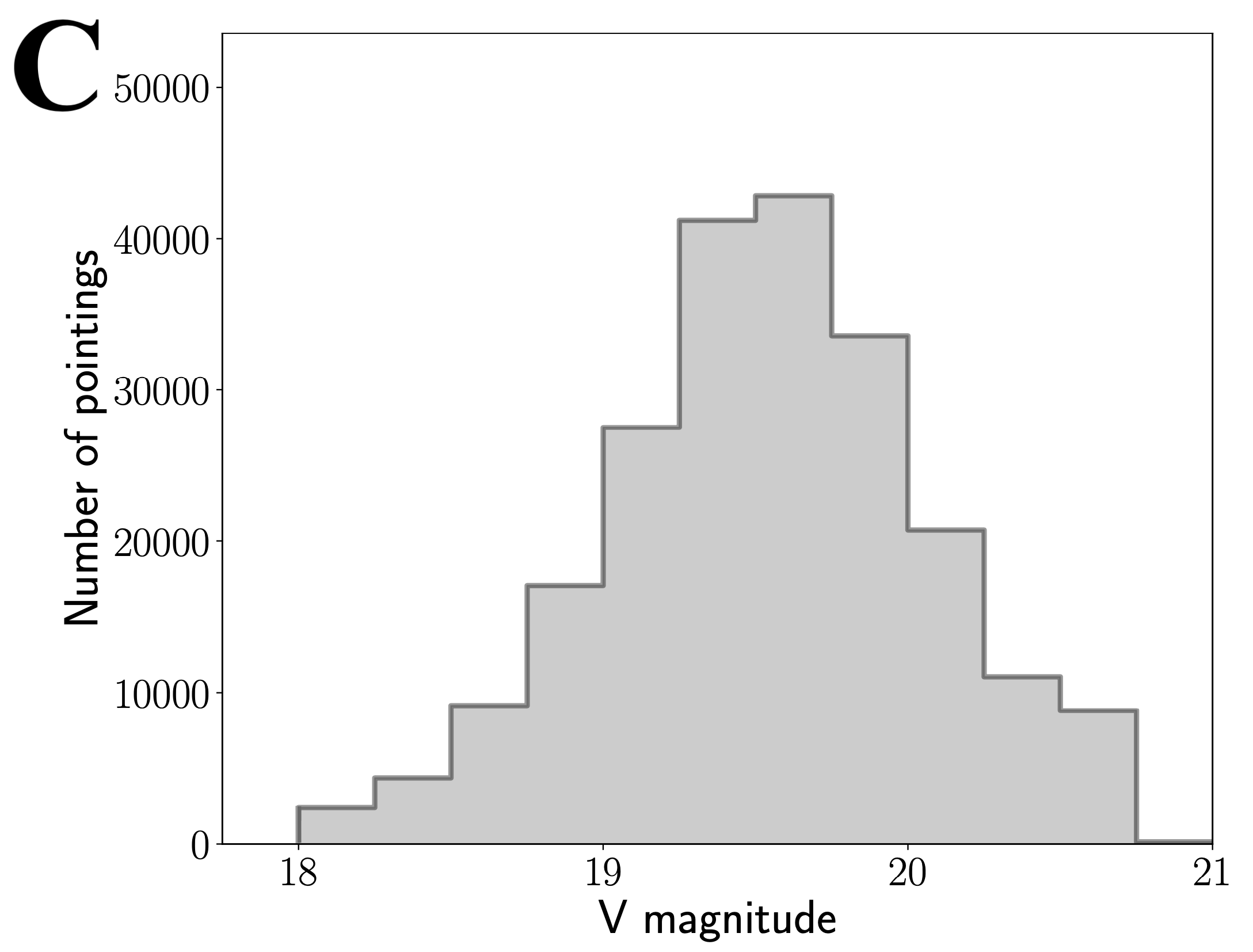}
\caption{$V$ magnitude and rate of motion distributions of synthetic inner-Venus asteroid detections. (A) Comparison of the $V$ magnitude distribution of synthetic \an asteroid detections (grey histogram) with the $V$ magnitude distribution weighted by the per detection efficiency as defined by Eq.~1 (red histogram). (B) Same as (A), but for the synthetic \anns's rate of motion. The vertical dashed lines in (A) and (B) are the values for \an on 2020 January 4. (C) The exposure-based 5-$\sigma$ limiting magnitude of the Twilight survey pointings between 2019 September 20 UTC and 2020 January 30 UTC with no assumptions or corrections regarding the efficiency of the detections in the pointings.}
\end{figure}

\end{document}